\documentclass[12pt]{article}
\def\theequation{\arabic{section}.\arabic{equation}}
\usepackage{amssymb}

\newcounter{rown}

\begin{document}
\renewcommand{\thefootnote}{\fnsymbol{footnote}}
\renewcommand{\theequation}{\thesection.\arabic{equation}}
\title{Superalgebras of (split-)division algebras and the split octonionic $M$-theory in
$(6,5)$-signature}
\author{Zhanna Kuznetsova\thanks{{\em e-mail: zhanna@cbpf.br}} 
~and Francesco Toppan\thanks{{\em e-mail: toppan@cbpf.br}}
\\ \\
{\it $~^\ast$ICE-UFJF, cep 36036-330, Juiz de Fora (MG), Brazil}
\\ 
{\it $~^\dagger$ CBPF, Rua Dr.}
{\it Xavier Sigaud 150,}
 \\ {\it cep 22290-180, Rio de Janeiro (RJ), Brazil}}
\maketitle
\begin{abstract}
The connection of (split-)division algebras with Clifford algebras and supersymmetry
is investigated. At first we introduce the class of superalgebras constructed from
any given (split-)division algebra. 
We further specify which real Clifford algebras and real fundamental spinors can be reexpressed
in terms of split-quaternions. Finally, we construct
generalized supersymmetries admitting bosonic tensorial central charges in terms
of (split-)division algebras. In particular we prove that split-octonions allow
to introduce a split-octonionic $M$-algebra which extends to the $(6,5)$ signature the properties of the $11$-dimensional octonionic $M$-algebras (which only exist in the $(10,1)$ Minkowskian and $(2,9)$ signatures).
\end{abstract}
\vfill 
\rightline{CBPF-NF-014/06}

\newpage
\section{Introduction}

The connection between division algebras and supersymmetry is well established since the \cite{kt} paper. Along the years,
this connection has been further clarified, see e.g. \cite{toppan}. It is also well-known that division algebras are at the core of the
classification of Clifford algebras and spinors, \cite{por} and \cite{oku}. The inter-relation of these mathematical
structures has played a major role in a vast set of physical applications, ranging from supergravity (as well as superstrings
and $M$-theory) to the construction of supersymmetric integrable models. Comparatively less is known, however,
when we consider the case of the split version of division algebras (see \cite{mcC} for an introduction to split-division algebras).
Split-division algebras have been used with profit both in mathematical applications like, e.g., the generalization of the Tits-Freudenthal
magic square construction to split division algebras \cite{bs}, as well as in more physically motivated applications, like the recent interesting reformulation of electrodynamics, see \cite{gog1} and \cite{gog2}, in terms of split-octonions.\par
The purpose of this paper is to clarify the interrelation between (split)-division algebras and the graded algebras
that can be obtained from them, as well as the role played by the split-division algebras in the construction of Clifford algebras and spinors. Finally, we will construct the generalized supersymmetries associated to them. As an extra bonus, we will be able to
solve a puzzle concerning the octonionic version of the $M$-theory, see \cite{lt1} and \cite{lt2}, proving the
existence of a split-octonionic $M$-algebra existing in the exotic $(6,5)$ signature (more on that later). \par
This paper is so conceived, at first we introduce the whole set of (split)-division algebras through the unified
framework provided by the (generalized) Cayley-Dickson doubling construction. We further point out which graded algebras can be obtained as (anti-)commutators algebras from a given split-division algebra composition law.
Tables will further be produced, extending to split-quaternions the results of \cite{por} concerning the division-algebra character of Clifford algebras and spinors. In given space-time signatures spinors which are valued either in the split-quaternions or in the split-octonions are produced. This allows to extend to split-quaternions or to split-octonions the construction of (constrained) generalized supersymmetries presented
in \cite{crt2}, \cite{top} and \cite{kuto}. The split-octonionic $M$-algebra, existing in the $(6,5)$ signature, is a particular example of this construction.
In the Conclusions we make further comments about the implications and the physical relevance of the results here obtained. For the moment we point out that higher-dimensional (generalized) supersymmetries formulated in space-time dimensions $D\geq 8$ admit the peculiar feature that they come in several related versions in given signatures.
The associated
supersymmetric theories are all dually related (``the space-time dualities" of ref. \cite{hul}). The $10$-dimensional superstrings, e.g., only exist
in three ($(9,1)$, $(1,9)$ and $(5,5)$) signatures, the latter presenting five time directions. The $11$-dimensional supergravities are encountered, besides
the Minkowskian $(10,1)$ signature, also in the exotic $(2,9)$ and $(6,5)$
signatures.
It was proven in \cite{drt} that such dually-related versions are in
consequence of the triality of the $D=8$ dimensions (the dually related theories are indeed triality related and close the $S_3$ group). 
We recall that it is eight-dimensional the transverse space of both the light-cone formulation of the $10$-dimensional superstrings and of the supermembranes evolving in a flat $11$-dimensional target spacetime. In $D=8$, the triality allowed
signatures are $(8,0)$, $(0,8)$ and the exotic $(4,4)$.
Both the original Cartan's triality, see \cite{bae}, and the space-time triality
of ref. \cite{drt} are a consequence of the octonions. On the other hand, 
it was quite puzzling that, while the standard $M$-algebra (based on real spinors) exists for the
whole set of above signatures, the octonionic $M$-algebra of ref. \cite{lt1} and \cite{lt2} only exists
in $(10,1)$ and $(2,9)$ (the $(6,5)$ signature is missing). The reason being essentially
due to the fact that the seven imaginary octonions have to be
accommodated either in the time-like or in the space-like directions. Obviously $7$ cannot enter either $6$ or $5$. 
By relaxing the condition of dealing with division-algebras, we can here solve the puzzle by expressing the
counterpart of the octonionic $M$-algebra in the exotic signature
$(6,5)$ in terms of the split-octonions. It is worth mentioning that the $(6,5)$ space-time also carries a supersymmetry
based on split-quaternionic spinors.\par 
The paper is structured as follows: in Section {\bf 2} we revisit the split-division algebras. In Section {\bf 3} we construct the (graded)-algebras associated to the (split)-division algebras. The tables relating Clifford algebras and spinors to 
split-quaternions are furnished in Section {\bf 4}. The generalized supersymmetries based on split-division algebras
and the split-octonionic $M$-algebra are presented in Section {\bf 5}. We produce here also the free actions
for split-quaternionic and split-octonionic spinors. In the Conclusions, we provide comments
on the results here obtained. In order 
to make the paper self-consistent we present in appendix the generalized Cayley-Dickson doubling.

\section{Split-division algebras revisited}

The construction of split-division algebras in terms of the Cayley-Dickson doubling procedure is reviewed in the Appendix. For later purposes it is useful to explicitly present here
the (split-)division algebras structure constants, conjugations and quadratic forms in 
the case of quaternions (${\Bbb H}$), split-quaternions (${\widetilde {\Bbb H}}$),
octonions (${\Bbb O}$) and split-octonions (${\widetilde {\Bbb O}}$).
Complex (${\Bbb C}$) and split-complex (${\widetilde {\Bbb C}}$) numbers are immediately
recovered as sub-algebra of, let's say, the split-quaternions.\par 
Let us introduce at first the quaternions.
The three imaginary quaternions $e_i\in{\Bbb H}$ ($i=1,2,3$) satisfy the relations
\begin{eqnarray}
e_i\cdot e_j&=&-\delta_{ij}{\bf 1} + \epsilon_{ijk} e_k
\end{eqnarray}
($\epsilon_{ijk}$ is the totally antisymmetric tensor, normalized s.t. $\epsilon_{123}=1$).\par
The conjugation and the quadratic form (norm) are respectively given by
\begin{eqnarray}
{e_i}^{\ast} &=& -e_i,\nonumber\\
N(e_i) &=& 1.
\end{eqnarray}

For what concerns the octonions, we can introduce them as as ${\Bbb O} = {\Bbb H}-$ (see Appendix). Therefore, the seven imaginary octonions $E_i$ are recovered through the positions
\begin{eqnarray}
E_i &=& (e_i,0)\nonumber\\
E_{3+i} &=& (0,e_i)\nonumber\\
E_7 &=& -(0,1).
\end{eqnarray}
They satisfy the relations
\begin{eqnarray}{\label{cijk}}
E_i\cdot E_j &=& -\delta_{ij}{\bf 1} + C_{ijk} E_k,
\end{eqnarray}
while their conjugation and quadratic form are respectively given by
\begin{eqnarray}
{E_i}^{\ast} &=& - E_i,\nonumber\\
N(E_i)&=& 1.
\end{eqnarray}
In the above (\ref{cijk}) formula the $C_{ijk}$'s are the totally antisymmetric
octonionic structure constants, non-vanishing only for the triples\footnote{The seven imaginary octonions can be associated to the points of the seven-dimensional projective Fano's plane, while the triples correspond to the seven lines of this plane, see \cite{bae} for details.} 
\begin{eqnarray}\label{triples}
&C_{123}=C_{147}=C_{165}=C_{246}=C_{257}=C_{354}=C_{367}=1.&
\end{eqnarray}
With a similar procedure the split octonions can be expressed through the identificaion
${\widetilde {\Bbb O}}= {\Bbb H}+$.
The seven imaginary split-octonions ${\widetilde E}_i$ are given, as before, by
\begin{eqnarray}
{\widetilde E}_i &=& (e_i,0)\nonumber\\
{\widetilde E}_{3+i} &=& (0,e_i)\nonumber\\
{\widetilde E}_7 &=& -(0,1).
\end{eqnarray}
They satisfy the relations
\begin{eqnarray}
{\widetilde E}_i\cdot {\widetilde E}_j &=& -\eta_{ij}{\bf 1} + 
C_{ijk}\eta_{kr} {\widetilde E}_r,
\end{eqnarray}
together with
\begin{eqnarray}
{{\widetilde E}_i}^{\ast} &=& -{\widetilde E}_i,\nonumber\\
N({\widetilde E}_i)&=& \eta_{ii}.
\end{eqnarray}
In the above formulas $\eta_{ij}$ denotes the diagonal matrix $(+++----)$ with three positive and four negative eigenvalues (normalized to $\pm1$).\par
The quaternionic subalgebra ${\Bbb H}$ of the split octonions is obtained by
restricting the imaginary split-octonions ${\widetilde E}_i$ to the values $i=1,2,3$.\par 
On the other hand, the split-quaternionic 
subalgebra ${\widetilde {\Bbb H}}$ is recovered by taking any subset of three elements lying in the six other lines
of the Fano's projective plane (namely, the triples $(147)$, $(165)$, $(246)$,
$(257)$, $(354)$ and $(367)$). \par
The split-quaternions subalgebra can be explicitly presented as follows, in terms of
the three generators ${\widetilde e}_i$ $(i=1,2,3)$,
\begin{eqnarray}\label{splitquaterniongenerators}
{\widetilde e}_i\cdot {\widetilde e}_j &=& -\eta_{ij}{\bf 1} + 
\epsilon_{ijk}\eta_{kr} {\widetilde e}_r,
\end{eqnarray}
with conjugation and quadratic form given by
\begin{eqnarray}
{{\widetilde e}_i}^{\ast} &=& -{\widetilde e}_i,\nonumber\\
N({\widetilde e}_i)&=& \eta_{ii}.
\end{eqnarray}
$\eta_{ij}$  is now the diagonal matrix $(--+)$.\par
The split quaternions admit a faithful representation in terms of $2\times 2$ real matrices
given by
\begin{eqnarray}
\tau_1 = \left(
\begin{array}{cc}
0&1\\
1&0	
\end{array} 
\right)
&& \tau_2 = \left(
\begin{array}{cc}
1&0\\
0&-1	
\end{array}
 \right)\nonumber\\
 \tau_A = \left(
\begin{array}{cc}
0&1\\
-1&0	
\end{array} 
\right)
&& {\bf 1}_2 = \left(
\begin{array}{cc}
1&0\\
0&1	
\end{array}
\right)
 \end{eqnarray}
The conjugate element of a generic split-quaternion $X\in {\widetilde {\Bbb H}}$ is represented by
\begin{eqnarray}
X^\ast &=& -\tau_A X^T\tau_A.
\end{eqnarray}
\section{Graded algebras from (split-)division algebras}

The multiplication ``$\cdot$" of a composition algebra ${\Bbb A}$
induces on ${\Bbb A}$ the structure of a (graded) algebra
${\Bbb A}\times {\Bbb A}\rightarrow {\Bbb A}$  of (anti)commutators.
Namely, for $a,b\in {\Bbb A}$, we can introduce the algebra of graded brackets defined through
\begin{eqnarray}
\relax [a,b \} &=& ab +(-1)^{\epsilon_a\epsilon_b} ba,
\end{eqnarray}
where $\epsilon_{a,b} \equiv 0,1 ~ mod ~2$ corresponds to the ${\Bbb Z}_2$ grading of
the generators $a,b$ respectively.
The (anti)commutator algebra is a (graded) Lie algebra if the multiplication is
associative. If the multiplication is alternative (see the Appendix), the (anti)commutator
algebra is a (graded) Malcev algebra (see \cite{crt1} for its definition).\par
The ${\Bbb Z}_2$ grading implies for ${\Bbb A}$ the decomposition 
${\Bbb A} = {\Bbb A}_0\oplus {\Bbb A}_1$ such that, for non-vanishing $[a,b\}$,
\begin{eqnarray}
deg ({[a,b\}}) &=& deg(a)+deg(b) \equiv \epsilon_a+\epsilon_b ~(mod~2)
\end{eqnarray} 
We can easily list the set of admissible ${\Bbb Z}_2$ gradings for each one of the 
four division algebras (the ${\Bbb R}$ case is trivial). As a corollary, this gives us the list
of the admissible superalgebras based on each division algebra. 
From the previous section results we know that the
split-division algebras structure constants are recovered, up to a normalization factor, from the structure constants of their corresponding division algebra. For this reason the list of the admissible ${\Bbb Z}_2$ gradings (and associated superalgebras)
of division algebras can also be regarded as the list of admissible ${\Bbb Z}_2$ gradings
(and associated superalgebras) 
of the split-division algebras .
The identity is necessarily an even (bosonic) element of the (super)algebra and corresponds
to a central term. The (split) imaginary numbers close graded subalgebras of dimension
$1$ (for ${\Bbb C}$ and ${\widetilde{\Bbb C}}$),  $3$ (for ${\Bbb H}$ and ${\widetilde{\Bbb H}}$)
and $7$ (for ${\Bbb O}$ and ${\widetilde{\Bbb O}}$).\par
It is worth noticing that we can regard the (anti)commutators algebras induced by the composition law as {\em abstract} (super)algebras. In particular this implies that the ${\Bbb Z}_2$ superalgebra grading does not necessarily coincide with a ${\Bbb Z}_2$ grading of the
composition law (which requires satisfying
$deg (ab) = deg(a)+deg(b) ~mod~2$). This point can be better illustrated with an explicit example. Let's take the three imaginary quaternions $e_i$'s. If we assign odd-grading (fermionic character)
to $e_1$ and $e_2$, then $e_3$, appearing on the r.h.s. of the multiplication $e_1\cdot e_2$, is necessarily even-graded (bosonic). On the other hand, the anticommutator $\{e_1,e_2\}$
is vanishing. As far as the anticommutators {\em alone} are concerned, we can consistently assign
odd-grading to $e_3$ as well. In the following we will denote as ``compatible" the restricted class of (super)algebras whose ${\Bbb Z}_2$ grading is
an acceptable ${\Bbb Z}_2$ grading for the composition law. \par
The admissible ${\Bbb Z}_2$ gradings are expressed by the following table
(the last column refers to the {\em compatible} superalgebras). We have
\begin{eqnarray}&\label{z2gradings}
\begin{tabular}{|c|c|c|c|}\hline
  $$&$bosons/fermions$&$(super)algebra$&$compatibility$\\ \hline
${\mathbb C } $, ${\widetilde {\mathbb C }}$&$1B+0F$&$yes$&$yes$\\ 
$ $&$0B+1F$&$yes$&$yes$\\ \hline
${\mathbb H} $, ${\widetilde {\mathbb H}}$&$3B+0F$&$yes$&$yes$\\ 
$ $&$2B+1F$&$no$&$-$\\ 
$  $&$1B+2F$&$yes$&$yes$\\ 
$ $&$0B+3F$&$yes$&$no$\\ \hline
${\mathbb O } $, ${\widetilde {\mathbb O}}$&$7B+0F$&$yes$&$yes$\\ 
$ $&$6B+1F$&$no$&$-$\\ 
$  $&$5B+2F$&$no$&$-$\\ 
$ $&$4B+3F$&$no$&$-$\\ 
$ $&$3B+4F^{(a)}$&$no$&$-$\\ 
$ $&$3B+4F^{(b)}$&$yes$&$yes$\\ 
$  $&$2B+5F$&$no$&$-$\\ 
$ $&$1B+6F$&$yes$&$no$\\ 
$ $&$0B+7F$&$yes$&$no$\\ \hline 
\end{tabular}&\nonumber\\
&&\end{eqnarray}
There are two distinguished $3B+4F$ cases. The second one $(b)$ corresponds to the
three bosonic elements lying on one of the seven lines corresponding to the triples
in (\ref{triples}). Without loss of generality, the three octonionic elements in the line can always be chosen as $E_1$, $E_2$ and $E_3$ (${\widetilde E}_1$, ${\widetilde E}_2$,
${\widetilde E}_3$ for split-octonions).  Without loss of generality the case ($a$)
can be obtained by taking the three bosonic elements as $E_1$, $E_2$, $E_4$
(${\widetilde E}_1$, ${\widetilde E}_2$,
${\widetilde E}_4$ for split-octonions, respectively).
There is no superalgebra associated to the case $(a)$, while a compatible superalgebra is found
in the $(b)$ case.

\section{(Split-)division algebras, Clifford algebras \\
and spinors}

It is well-known that the Clifford algebras are related 
to the ${\Bbb R}$, ${\Bbb C}$, ${\Bbb H}$ associative
division algebras. The $Cl(s,t)$ Clifford algebra is defined as the enveloping algebra
generated by the gamma-matrices satisfying the relation
\begin{eqnarray}\label{cliffordgenerator}
\Gamma_i\Gamma_j+\Gamma_j\Gamma_i &=& 2\eta_{ij},
\end{eqnarray}
with $\eta_{ij}$ a diagonal matrix of $(s,t)$ signature
(i.e. $s$ positive, $+1$, and $t$ negative, $-1$, diagonal
entries, with $s$ and $t$ denoting, respectively, the number of space-like and time-like dimensions).
\par
 The most general irreducible {\em real}
matrix representation of a Clifford algebra is classified according to the property of the most
general $S$ matrix commuting with all the $\Gamma$'s ($\relax
[S,\Gamma_i ] =0$ for all $i$). If the most general $S$ is a
multiple of the identity, we get the normal (${\Bbb R}$) case.
Otherwise, $S$ can be the sum of two matrices, the second one
multiple of the square root of $-1$ (this is the almost complex,
${\Bbb C}$ case) or the linear combination of $4$ matrices closing
the quaternionic algebra (this is the ${\Bbb H}$ case). We obtain, for $s,t\leq 8$,
the following table, see \cite{por}
{\begin{eqnarray} &
\begin{tabular}{|c|c|c|c|c|c|c|c|c|c|}
\hline
   $s \backslash t $ &$0$& $1$&$2$&  $3$&$4$& $5$&$6$&$7$&$8$
      \\ \hline
      $0 $ &${\Bbb R}$& ${\Bbb C}$&${\Bbb H}$&  $~^2{\Bbb H}$&${\Bbb H}(2)$& ${\Bbb C}(4)$&${\Bbb R}(8)$&$~^2{\Bbb R}(8)$&${\Bbb R}(16)$
      \\ \hline
      $1$ &$~^2{\Bbb R}$& ${\Bbb R}(2)$&${\Bbb C}(2)$&  ${\Bbb H}(2)$&$~^2{\Bbb H}(2)$& ${\Bbb H}(4)$&${\Bbb C}(8)$&${\Bbb R}(16)$&$~^2{\Bbb R}(16)$
      \\ \hline
      $2 $ &${\Bbb R}(2)$& $~^2{\Bbb R}(2)$&${\Bbb R}(4)$&  ${\Bbb C}(4)$&${\Bbb H}(4)$& $~^2{\Bbb H}(4)$&${\Bbb H}(8)$&${\Bbb C}(16)$&${\Bbb R}(32)$
      \\ \hline
      
      $3 $ &${\Bbb C} (2)$& ${\Bbb R}(4)$&$~^2{\Bbb R}(4)$&  ${\Bbb R}(8)$&${\Bbb C}(8)$& ${\Bbb H}(8)$&$~^2{\Bbb H}(8)$&${\Bbb H}(16)$&${\Bbb C}(32)$
      \\ \hline
      $4$ &${\Bbb H}(2)$& ${\Bbb C}(4)$&${\Bbb R}(8)$&  $~^2{\Bbb R}(8)$&${\Bbb R}(16)$& ${\Bbb C}(16)$&${\Bbb H}(16)$&$~^2{\Bbb H}(16)$&${\Bbb H}(32)$
      \\ \hline
      $5 $ &$~^2{\Bbb H}(2)$& ${\Bbb H}(4)$&${\Bbb C}(8)$&  ${\Bbb R}(16)$&$~^2{\Bbb R}(16)$& ${\Bbb R}(32)$&${\Bbb C}(32)$&${\Bbb H}(32)$&$~^2{\Bbb H}(32)$
      \\ \hline
      $6 $ &${\Bbb H}(4)$& $~^2{\Bbb H}(4)$&${\Bbb H}(8)$&  ${\Bbb C}(16)$&${\Bbb R}(32)$& $~^2{\Bbb R}(32)$&${\Bbb R}(64)$&${\Bbb C}(64)$&${\Bbb H}(64)$
      \\ \hline
      $7 $ &${\Bbb C}(8)$& ${\Bbb H}(8)$&$~^2{\Bbb H}(8)$&  ${\Bbb H}(16)$&${\Bbb C}(32)$& ${\Bbb R}(64)$&$~^2{\Bbb R}(64)$&${\Bbb R}(128)$&${\Bbb C}(128)$
      \\ \hline
      $8 $ &${\Bbb R}(16)$& ${\Bbb C}(16)$&${\Bbb H}(16)$&  $~^2{\Bbb H}(16)$&${\Bbb H}(32)$& ${\Bbb C}(64)$&${\Bbb R}(128)$&$~^2{\Bbb R}(128)$&${\Bbb R}(256)$
      \\ \hline
\end{tabular}&\nonumber
\end{eqnarray}}
The famous $mod\quad 8$ property of Clifford algebras allows to extend the table above for values $s,t>8$.\par
The suffix ``$~^2$" in the $s-t= 1~mod~8$ entries is introduced to take into account that, for such coupled values of $s,t$, a faithful representation of the Clifford algebra is obtained as a direct sum of its two inequivalent irreducible representations,
see \cite{por} for details.\par
Following \cite{crt2} we have another possibility of understanding the connection
between Clifford algebras and division algebras.  
We can simply state that a Clifford algebra is of ${\Bbb R}$, ${\Bbb C}$ or ${\Bbb H}$ type if its fundamental
irreducible representation is realized in terms of matrices with entries in the corresponding division algebra.
A constructive way of proving the above statement makes use of the two lifting algorithms \cite{crt2}, expressing the $Cl(s+1,t+1)$ and $Cl(t+2,s)$ Clifford irreps in terms of the Clifford irreps of $Cl(s,t)$ (given by
the $s+t$ gamma $d\times d$ matrices $\gamma_i$'s). The $s+t+2$ gamma matrices  $\Gamma_j$ are given, in the two cases, by
\begin{eqnarray}
 \Gamma_i &\equiv& \left(
\begin{array}{cc}
  0& \gamma_i \\
  \gamma_i & 0
\end{array}\right), \quad \left( \begin{array}{cc}
  0 & {\bf 1}_d \\
  -{\bf 1}_d & 0
\end{array}\right),\quad \left( \begin{array}{cc}
  {\bf 1}_d & 0\\
  0 & -{\bf 1}_d
\end{array}\right)\nonumber\\
&&\label{one}
\end{eqnarray}
and, respectively,
\begin{eqnarray}
 \Gamma_j &\equiv& \left(
\begin{array}{cc}
  0& \gamma_i \\
  -\gamma_i & 0
\end{array}\right), \quad \left( \begin{array}{cc}
  0 & {\bf 1}_d \\
  {\bf 1}_d & 0
\end{array}\right),\quad \left( \begin{array}{cc}
  {\bf 1}_d & 0\\
  0 & -{\bf 1}_d
\end{array}\right)\nonumber\\
&&\label{two}
\end{eqnarray}
The spinors carry a representation of the $Spin(s,t)$ spin group (see \cite{por}),
whose Lie algebra generators are given by the gamma matrices commutators. As a result,
the division algebra
structure of Gamma matrices extends to spinors. There is, however, for certain signatures of the
space-time, a mismatch between division-algebra properties of the fundamental spinors
and their associated Clifford algebras, see \cite{lt1} and \cite{dflv}. The mismatch is due to the existence
of a Weyl-projection. We recall that, following \cite{crt2}, the fundamental spinors belong to the representation
of the spin group admitting maximal division algebra structure. A table, presenting the division-algebra
properties of spinors for $s,t\leq 8$, is here produced 
{\begin{eqnarray} &
\begin{tabular}{|c|c|c|c|c|c|c|c|c|c|}
\hline
   $s \backslash t $ &$0$& $1$&$2$&  $3$&$4$& $5$&$6$&$7$&$8$
      \\ \hline
      $0 $ && ${\Bbb R}^W$&${\Bbb C}^W$&  ${\Bbb H}$&${\Bbb H}^W$& ${\Bbb H}(2)^W$&${\Bbb C}(4)^W$&${\Bbb R}(8)$&${\Bbb R}(8)^W$
      \\ \hline
      $1$ &${\Bbb R}$& ${\Bbb R}^W$&${\Bbb R}(2)^W$&  ${\Bbb C}(2)^W$&${\Bbb H}(2)$& ${\Bbb H}(2)^W$&${\Bbb H}(4)^W$&${\Bbb C}(8)^W$&${\Bbb R}(16)$
      \\ \hline
      $2 $ &${\Bbb C}^W$& ${\Bbb R}(2)$&${\Bbb R}(2)^W$&  ${\Bbb R}(4)^W$&${\Bbb C}(4)^W$& ${\Bbb H}(4)$&${\Bbb H}(4)^W$&${\Bbb H}(8)^W$&${\Bbb C}(16)^W$
      \\ \hline
      
      $3 $ &${\Bbb H}^W$& ${\Bbb C}(2)^W$&${\Bbb R}(4)$&  ${\Bbb R}(4)^W$&${\Bbb R}(8)^W$& ${\Bbb C}(8)^W$&${\Bbb H}(8)$&${\Bbb H}(8)^W$&${\Bbb H}(16)^W$
      \\ \hline
      $4$ &${\Bbb H}^W$& ${\Bbb H}(2)^W$&${\Bbb C}(4)^W$&  ${\Bbb R}(8)$&${\Bbb R}(8)^W$& ${\Bbb R}(16)^W$&${\Bbb C}(16)^W$&${\Bbb H}(16)$&${\Bbb H}(16)^W$
      \\ \hline
      $5 $ &${\Bbb H}(2)$& ${\Bbb H}(2)^W$&${\Bbb H}(4)^W$&  ${\Bbb C}(8)^W$&${\Bbb R}(16)$& ${\Bbb R}(16)^W$&${\Bbb R}(32)^W$&${\Bbb C}(32)^W$&${\Bbb H}(32)$
      \\ \hline
      $6 $ &${\Bbb C}(4)^W$& ${\Bbb H}(4)$&${\Bbb H}(4)^W$&  ${\Bbb H}(8)^W$&${\Bbb C}(16)^W$& ${\Bbb R}(32)$&${\Bbb R}(32)^W$&${\Bbb R}(64)^W$&${\Bbb C}(64)^W$
      \\ \hline
      $7 $ &${\Bbb R}(8)^W$& ${\Bbb C}(8)^W$&${\Bbb H}(8)$&  ${\Bbb H}(8)^W$&${\Bbb H}(16)^W$& ${\Bbb C}(32)^W$&${\Bbb R}(64)$&${\Bbb R}(64)^W$&${\Bbb R}(128)^W$
      \\ \hline
      $8 $ &${\Bbb R}(8)^W$& ${\Bbb R}(16)^W$&${\Bbb C}(16)^W$&  ${\Bbb H}(16)$&${\Bbb H}(16)^W$& ${\Bbb H}(32)^W$&${\Bbb C}(64)^W$&${\Bbb R}(128)$&${\Bbb R}(128)^W$
      \\ \hline
\end{tabular}&\nonumber
\end{eqnarray}}
The ``$W$" denotes the presence of the Weyl projection. The numbers denote the dimensionality of the spinors.
Just like the previous table, the division algebra properties of fundamental spinors for $s,t>8$ are recovered from the $mod~8$ property of Clifford algebras.\par
The same type of analysis leading to the division-algebra properties of, respectively, Clifford algebras and fundamental spinors, can be repeated when investigating split-division algebra properties.
The interesting case is that of split-quaternions (${\widetilde{\Bbb H}}$) since, unlike the division-algebra
case, split complex numbers and split quaternions are both represented in terms of $2\times 2$ real matrices
(complex numbers are represented by two $2\times 2$ real matrices and quaternions by $4\times 4$ real matrices).
The basic example is provided by the $Cl(2,1)$ Clifford algebra, whose fundamental relation (\ref{cliffordgenerator})
can be realized in terms of the three split-quaternions of (\ref{splitquaterniongenerators}). The application of the lifting algorithms (\ref{one})
and (\ref{two}) allows to induce a split-quaternionic structure for the
$Cl(s,t)$ Clifford algebras with
\begin{eqnarray}\label{stone}
s= 2+k &,& t=1+8m+k, \quad for\quad m,k=0,1,2,\ldots 
\end{eqnarray}
and
\begin{eqnarray}\label{sttwo}
s= 3+8m+k &,& t=2+k, \quad for\quad m,k=0,1,2,\ldots
\end{eqnarray}
These Clifford algebras are the ``oxidized forms" (according to \cite{kuto}). In analogy with the construction in \cite{crt2}, reduced split-quaternionic
Clifford algebras are obtained for $Cl(s-1,t)$ and $Cl(s-2,t)$, where $s,t$ are either given by
(\ref{stone}) or by (\ref{sttwo}).\par
At the end we obtain the table of split-quaternionic Clifford algebras given by   
{\begin{eqnarray} &
\begin{tabular}{|c|c|c|c|c|c|c|c|c|}
\hline
   $s \backslash t $&  $1$&$2$&  $3$&$4$& $5$&$6$&$7$&$8$
      \\ \hline
      $1$& ${\widetilde{\Bbb H}}$&$0$&  $0$&$0$& $0$&$0$&$0$&$0$
      \\ \hline
      $2  $&$~^2{\widetilde{\Bbb H}}$&${\widetilde{\Bbb H}}(2)$&  $0$&$0$& $0$&$0$&$0$&${\widetilde{\Bbb H}}(16)$
      \\ \hline
      
      $3 $ & ${\widetilde{\Bbb H}}(2)$&$~^2{\widetilde{\Bbb H}}(2)$&  ${\widetilde{\Bbb H}}(4)$&$0$& $0$&$0$&$0$&$0$
      \\ \hline
      $4$ & $0$&${\widetilde{\Bbb H}}(4)$&  $~^2{\widetilde{\Bbb H}}(4)$&${\widetilde{\Bbb H}}(8)$& $0$&$0$&$0$&$0$
      \\ \hline
      $5 $ & $0$&$0$&  ${\widetilde{\Bbb H}}(8)$&$~^2{\widetilde{\Bbb H}}(8)$& ${\widetilde{\Bbb H}}(16)$&$0$&$0$&$0$
      \\ \hline
      $6 $ & $0$&$0$&  $0$&${\widetilde{\Bbb H}}(16)$& $~^2{\widetilde{\Bbb H}}(16)$&${\widetilde{\Bbb H}}(32)$&$0$&$0$
      \\ \hline
      $7 $ & $0$&$0$&  $$&$0$& ${\widetilde{\Bbb H}}(32)$&$~^2{\widetilde{\Bbb H}}(32)$&${\widetilde{\Bbb H}}(64)$&$0$
      \\ \hline
      $8 $ & $0$&$0$&  $0$&$0$& $0$&${\widetilde{\Bbb H}}(64)$&$~^2{\widetilde{\Bbb H}}(64)$&${\widetilde{\Bbb H}}(128)$
      \\ \hline
\end{tabular}&\nonumber
\end{eqnarray}}
Similarly, the split-quaternionic table for fundamental spinors is given by
{\begin{eqnarray} &
\begin{tabular}{|c|c|c|c|c|c|c|c|c|}
\hline
   $s \backslash t $& $1$&$2$&  $3$&$4$& $5$&$6$&$7$&$8$\\ \hline
      $1$ &0& ${\widetilde{\Bbb H}}^W$&$0$&  $0$&$0$& $0$&$0$&$0$
      \\ \hline
      $2 $ &${\widetilde{\Bbb H}}$& ${\widetilde{\Bbb H}}^W$&${\widetilde{\Bbb H}}(2)^W$&  $0$&$0$& $0$&$0$&$0$
      \\ \hline
      
      $3 $ &$0$& ${\widetilde{\Bbb H}}(2)$&${\widetilde{\Bbb H}}(2)^W$&  ${\widetilde{\Bbb H}}(4)^W$&$0$& $0$&$0$&$0$
      \\ \hline
      $4$ &$0$& $0$&${\widetilde{\Bbb H}}(4)$&  ${\widetilde{\Bbb H}}(4)^W$&${\widetilde{\Bbb H}}(8)^W$& $0$&$0$&$0$
      \\ \hline
      $5 $ &$0$& $0$&$0$&  ${\widetilde{\Bbb H}}(8)$&${\widetilde{\Bbb H}}(8)^W$& ${\widetilde{\Bbb H}}(16)^W$&$0$&$0$
      \\ \hline
      $6 $ &$0$& $0$&$0$&  $0$&${\widetilde{\Bbb H}}(16)$& ${\widetilde{\Bbb H}}(16)^W$&${\widetilde{\Bbb H}}(32)^W$&$0
      $
      \\ \hline
      $7 $ &$0$& $0$&$0$&  $0$&$0$& ${\widetilde{\Bbb H}}(32)$&${\widetilde{\Bbb H}}(32)^W$&${\widetilde{\Bbb H}}(64)^W$
      \\ \hline
      $8 $ &$0$& $0$&$0$&  $0$&$0$& $0$&${\widetilde{\Bbb H}}(64)$&${\widetilde{\Bbb H}}(64)^W$
      \\ \hline
\end{tabular}&\nonumber
\end{eqnarray}}
Both tables above can be extended for $s,t>8$ due to the $mod~8$ property of Clifford algebras.

\section{Split-division algebras and generalized supersymmetries}

In this Section we discuss a physical application of both split-quaternions and split-octonions.
Essentially, we will prove that the constructions concerning quaternionic and octonionic spinors,
carried out in \cite{crt2}, can be extended to the split-quaternionic and the split-octonionic cases.
The main results in \cite{crt2} include {\em i}) the construction of free invariant actions for 
quaternionic and octonionic spinors and {\em ii}) the construction of quaternionic and octonionic
generalized supersymmetries. We explicitly discuss here which modifications have to be introduced in the
split cases w.r.t. their non split counterparts.\par
The notion of octonionic and split-octonionic spinors requires the introduction of the
(split-)octonionic realizations of the (\ref{cliffordgenerator}) relation, in terms of matrices with (split-)octonionic entries. The meaning of the octonionic realizations of (\ref{cliffordgenerator}) has been fully
described in \cite{crt2}. The same considerations apply to the split-octonionic cases as well.
The results furnished in this paper will concern the so-called {\em maximal Clifford algebras} leading to the
{\em oxidized supersymmetries}, see \cite{kuto} for a definition. Basically, the maximal Clifford algebras correspond
to the space-times of maximal dimension
supporting spinors of a given size. The results for the non-maximal space-times are simply obtained via a dimensional reduction
of the oxidized cases by applying the
formulas produced in \cite{kuto} (see this reference for a full discussion).\par
A feature distinguishing the split-quaternionic and octonionic cases w.r.t. the non-split ones is the fact that,
in the split case, the same space-time of a given signature can carry both split-quaternionic and split-octonionic spinors. In the non-split case, quaternionic and octonionic spinors are carried by space-times
of different signatures. For instance, in $D=11$ dimensions, the octonionic spinors can be introduced in the Minkowskian $(10,1)$ signature, while quaternionic spinors are associated with the Euclidean $(0,11)$ space-time. On the other hand, the $(6,5)$ signature carries both split-quaternionic and split-octonionic spinors. 
The ``oxidized" split-quaternionic space-times are given by formulas (\ref{stone}) and (\ref{sttwo}), while
the oxidized split-octonionic space-times are restricted by the conditions
\begin{eqnarray}\label{splitoctone}
s= 4+k &,& t=3+8m+k, \quad for\quad m,k=0,1,2,\ldots 
\end{eqnarray}
and
\begin{eqnarray}\label{splitocttwo}
s= 5+8m+k &,& t=4+k, \quad for\quad m,k=0,1,2,\ldots
\end{eqnarray}
We further point out that the $D$ generating gamma-matrices entering (\ref{cliffordgenerator}) for a $D$-dimensional space-time
are given, in the split-quaternionic case, by $D-3$ purely real matrices, while the remaining three matrices are given by the three split-imaginary split-quaternions multiplying the same purely real matrix (let's call it ``$T$"). In the split-octonionic case they are given by $D-7$ purely real
matrices plus the seven matrices obtained by the seven split-imaginary split-octonions multiplying a single, purely real matrix $T$.\par
We need at first to introduce the two matrices, usually denoted as $C$ and $A$ in the literature (see \cite{crt2}), which are related to the transposition and the hermitian conjugation respectively. $C$ is the
charge-conjugation matrix. For the maximal Clifford algebras it is uniquely defined and is given, up to a sign
factor, by the product of all symmetric (or all antisymmetric) generating gamma matrices. The matrix $A$ in the
split-quaternionic and split-octonionic cases can be defined as follows. If the purely real matrix $T$ introduced above is symmetric, then $A$ is given by the product of the subset of purely real gamma matrices of space-like type. Conversely, if $T$ is antisymmetric, $A$ is the product of the subset of purely real gamma matrices of time-like type.\par
The importance of a non-trivial conjugation is reflected by the fact that the supersymmetry algebra can be decomposed
in the three relations below. We have, for split-quaternionic or split-octonionic supercharges $Q_a$, the following relations
\begin{eqnarray}\label{superalgebra}
\{Q_a,Q_b\} = W_{ab}, &\quad& \{ {Q_a}^{\ast}, {Q_b}^\ast\} = {W_{ab}}^\ast, \nonumber\\
& \{Q_a, {Q_b}^\ast\} = Z_{ab}.
\end{eqnarray}    
The r.h.s. matrices $W_{ab}$ and $Z_{ab}$ contain the bosonic degrees of freedom of the superalgebra.
$W_{ab}$ is a symmetric matrix, while $Z_{ab}$ is hermitian. The bosonic r.h.s. can be expanded in terms
of the antisymmetrized products of the gamma matrices, namely
$W_{ab} = \sum_k (C\Gamma_{[\mu_1\ldots \mu_k]})_{ab} W^{[\mu_1\ldots\mu_k]}$ and
$Z_{ab} =\sum_k (A\Gamma_{[\mu_1\ldots\mu_k]})_{ab} Z^{[\mu_1\ldots\mu_k]}$, with the sum over $k$
restricted to symmetric or hermitian matrices.  
For split-quaternions and split-octonions, just like their quaternionic and octonionic counterparts
(and unlike the real and complex cases) the decomposition of the symmetric matrix $W_{ab}$ has to be limited
to, at most, $k=0$ and $k=1$. The reasons discussed in \cite{crt2} apply to the split-cases as well. No such
a limitation exists for the hermitian $Z_{ab}$ matrices.  The admissible integers $k$ label the higher-rank
tensors sectors of the bosonic r.h.s. . These sectors will be compactly denoted as ``$M_k$". \par
Due to the non-associativity, in the split-octonionic case (similarly as for the octonionic case), it must be
consistently specified the order in taking the antisymmetric product of $k>2$ (split-)octonionic valued gamma matrices. The correct prescription, given in \cite{crt2}, applies also to the split-octonions. It is given by
the formula 
\begin{eqnarray}
\relax [\Gamma_{1}\cdot \Gamma_{2}\cdot \dots \cdot \Gamma_k]
&\equiv& \frac{1}{k!}\sum_{perm.} (-1)^{\epsilon_{i_1\dots i_k}}
(\Gamma_{i_1}\cdot \Gamma_{i_2}\dots \cdot \Gamma_{i_k}),
\label{antisym}
\end{eqnarray}
where $(\Gamma_1\cdot \Gamma_2\dots \cdot \Gamma_k)$ denotes the
symmetric product
\begin{eqnarray}
(\Gamma_1\cdot \Gamma_2 \cdot\dots  \cdot \Gamma_k) &\equiv&
\frac{1}{2}(. ((\Gamma_1 \Gamma_2)\Gamma_3\dots )\Gamma_k)
+\frac{1}{2} (\Gamma_1(\Gamma_2(\dots \Gamma_k)).).
\end{eqnarray}
According to the discussion of \cite{top} (see also \cite{kuto}), the generalized superalgebras (\ref{superalgebra}) are divided into classes, in accordance to whether the bosonic degrees of freedom enter
either the symmetric matrix $W_{ab}$ or the hermitian matrix $Z_{ab}$. In application to our
split-cases, we obtain the following list of generalized superalgebras:\par
{\em i}) the split-quaternionic symmetric case,  
{\em ii}) the split-octonionic symmetric case, {\em iii}) the split-quaternionic hermitian case and,
finally, {\em iv}) the split-octonionic hermitian case.\par
The bosonic sectors in the four cases above are given by the following tables. 
For each admissible split-quaternionic or split-octonionic space-time signature of total dimension $D$ we provide the decomposition of the bosonic sector into rank-$k$ totally antisymmetric tensors and the total number of the associated bosonic degrees of freedom. We have
\par
 {\em i}) Split-quaternionic symmetric case,
 
 {{\begin{eqnarray}&
\begin{tabular}{|c|c|c|}\hline

$D=3$& $M0+M1$& $1+3$\\ \hline

$D=5$&$M1$&$5$\\ \hline

$D=7$&$-$&$-$\\ \hline

$D=9$&$M0$&$1$\\ \hline

$D=11$&$M0+M1$&$1+11$\\ \hline

$D=13$&$M1$&$13$\\ \hline

\end{tabular}&\label{tablem}\end{eqnarray}}}
 
 {\em ii}) Split-octonionic symmetric case,
 
  {{\begin{eqnarray}&
\begin{tabular}{|c|c|c|}\hline

$D=7$& $M0+M1$& $1+7$\\ \hline

$D=9$&$M1$&$9$\\ \hline

$D=11$&$-$&$-$\\ \hline

$D=13$&$M0$&$1$\\ \hline

\end{tabular}&\label{tablem2}\end{eqnarray}}}
 
 {\em iii}) Split-quaternionic hermitian case,
 
  {{\begin{eqnarray}&
\begin{tabular}{|c|c|c|}\hline

$D=3$& $M0$& $1$\\ \hline

$D=5$&$M0+M1$&$1+5$\\ \hline

$D=7$&$M1+M2$&$7+21$\\ \hline

$D=9$&$M2+M3$&$36+84$\\ \hline

$D=11$&$M0+M3+M4$&$1+165+330$\\ \hline

$D=13$&$M0+M1+M4+M5$&$1+13+715+1287$\\ \hline

\end{tabular}&\label{tablem3}\end{eqnarray}}}

  {\em iv}) Split-octonionic hermitian case,

  {{\begin{eqnarray}&
\begin{tabular}{|c|c|c|}\hline

$D=7$&$M0$&$1$\\ \hline

$D=9$&$M0+M1\equiv M4$&$1+9=10$\\ \hline

$D=11$&$M1+M2\equiv M5$&$11+41=52$\\ \hline

$D=13$&$M2+M3\equiv M6$&$64+168=232$\\ \hline

\end{tabular}&\label{tablem4}\end{eqnarray}}}

Some comments are in order. The split-octonionic hermitian supersymmetry implies an equivalence
among the rank-$k$ totally antisymmetric sectors. The maximal number of $52$ bosonic degrees of freedom
in the $(6,5)$ split-octonionic space-time can be described either as a single rank $k=5$ totally antisymmetric tensor, or 
as a combination of the rank $k=1$ and $k=2$ totally antisymmetric tensors. This feature is peculiar
to the (split-)octonionic supersymmetry and is in consequence of the non-associativity of the
(split-)octonions. For the $(6,5)$ spacetime no symmetric (split-)octonionic supersymmetry exists, so that
the maximal number of bosonic degrees of freedom necessarily enter the hermitian matrix $Z_{ab}$. \par
The same space-time carries also (split-)quaternionic spinors. A supersymmetric algebra involving split-quaternions for the $(6,5)$ spacetime implies at most $496$ bosonic degrees of freedom coming from
the hermitian sector, plus at most $1+11=12$ bosonic degrees of freedom associated with the symmetric
(i.e. the $W_{ab}$ matrix) sector. We should remind that the most general supersymmetry in the $(6,5)$ space-time based on {\em real} spinors contains at most $528$ bosonic degrees of freedom (the symmetric entries
of a $32\times 32$ real matrix). Summarizing, in the $(6,5)$ space-time example we can introduce
three different types of supersymmetries, in accordance with the choice of the fundamental spinors. We can indeed have
real, split-quaternionic or split-octonionic supersymmetries associated to this spacetime. The properties of
each type of supersymmetry are in consequence of the original choice of the basic spinors (real, split-quaternionic
or split-octonionic).  \par 
Let us conclude this section producing the formulas for the free action  of the
split-quaternionic and split-octonionic spinors $\Psi$. Following \cite{crt2} we can express the lagrangian
${\cal L}$ as a sum ${\cal L} = {\cal K} + {\cal M}$, where the kinetic term ${\cal K}$ and the massive
term ${\cal M}$ are given by  
\begin{eqnarray}\label{freeaction}
 {\cal K} &=& \frac{1}{2}tr[(\Psi^{\dagger} A
\Gamma^{\mu})\partial_{\mu} \Psi]+
\frac{1}{2}tr[\Psi^{\dagger} (A \Gamma^{\mu}\partial_{\mu}
\Psi)], \nonumber\\
{\cal M} &=& tr(\Psi^{\dagger}A \Psi)
\end{eqnarray}
In the above formulas the ``trace" {\em tr} refers to the projection over the identity for elements
of the (split-)division algebra (i.e. $tr( x_0+x_je_j)=x_0$). The brackets are inserted to take care of the
correct ordering when dealing with split-octonions, due to their non-associativity. The formula
(\ref{freeaction}) clearly holds also for the split-quaternionic case.\par
One can easily prove that no massive terms are allowed in the split-quaternionic space-times $(2,1)$
or $(3,2)$. Similarly, no massive terms are allowed for the $(4,3)$ and $(5,4)$ split-octonionic spacetimes.
The most interesting spacetime for the connection with the $M$-theory and its triality properties, as
discussed
in the Introduction, is $(6,5)$. It allows a non-vanishing mass-term for split-octonionic spinors.
\section{Conclusions}
In this work we have analyzed the connection of (split-)division algebras with Clifford algebras,
spinors and supersymmetry. More specifically, after having reviewed the generalized Cayley-Dickson 
double construction which allows to introduce, in a unified framework, all seven inequivalent 
(split-)division algebras, we analyzed the following points. At first we derived the (graded) algebras
of (anti)commutators which can be constructed from the (split-)division algebras composition law. Next,
we produced the tables of Clifford algebras and fundamental spinors related to the split-quaternions.\par
In Section {\bf 5} we made use of the conjugation properties of the split-division algebras in order to
construct, in the case of space-time signatures supporting either split-quaternionic or split-octonionic
valued spinors, the corresponding (constrained) generalized supersymmetries.
We also produced the free lagrangians for both the split-quaternionic and the split-octonionic spinors.
Concerning supersymmetries, we pointed out that in several examples the same space-time can carry both
split-quaternionic and split-octonionic supersymmetries. 
 The most interesting application
concerned the construction of the split-octonionic $M$-algebra, available in the exotic $(6,5)$-signature.
The construction of split-octonionic generalized supersymmetries parallels the construction, already available
in the literature (\cite{{lt1},{lt2},{crt2}}), of the octonionic generalized supersymmetries.\par
As recalled in the Introduction, higher-dimensional (for $D\geq 8$) supersymmetries hide a fundamental ambiguity,
due to the existence of triality-related formulations in different signatures.
This feature is exemplified, for instance, in the case of superstrings. The criticality condition allows to
determine the overall dimension ($=10$) of the target spacetime, but not its signature, which could be
either $(9,1)$, $(1,9)$ or the exotic $(5,5)$. A mathematically consistent superstring theory can be formulated
not only for a Minkowski target, but also in a $(5,5)$ spacetime. In \cite{hul} the web of dualities of the
corresponding theories (and their consequences for the dimensionally reduced theories) were explored.\par
So far this construction only existed for theories formulated in terms of real-valued (Majorana or Majorana-Weyl)
spinors, but not for their octonionic counterparts. The introduction of split-octonionic supersymmetries
in exotic signatures allows to extend the web of dualities to the (split-)octonionic formulations.
It immmediately implies, e.g., the existence (for the exotic $(5,5)$ signature) of a split-octonionic version of the \cite{chsu} octonionic formulation of the superstrings.
In the case of generalized supersymmetries with tensorial central charges, we obtain the split-octonionic version 
of the $M$-algebra in $(6,5)$ signature, sharing the same features as its octonic counterpart
(the most noticeable property being the dependence of the rank-$5$ totally antisymmetric tensors in terms of a combination of the rank-$1$ and rank-$2$ antisymmetric tensors). The $11$-dimensional $M$-algebra admits an equivalent
$12$-dimensional presentation, the $F$-algebra formulation \cite{crt2} which, in case of the split-octonions, is 
realized in the
$(6,6)$ signature. It is worth reminding that it has been suggested, see e.g. \cite{boya} and references therein, that since 
octonions are at the very core of many mathematical exceptional structures, a possible exceptional formulation
for a ``Theory Of Everything" would require an octonionic formulation. If this is indeed the case, it would be
expected that the octonionic and split-octonionic versions of the $M$-algebra should play a major role.

\renewcommand{\theequation}{A.\arabic{equation}}
\setcounter{equation}{0}
\par{~}\\
{\Large{\bf Appendix }}\\{~}\\

We collect here for convenience, following \cite{mcC} and \cite{bs}, 
the main properties and definition of (split-)division algebras. \par 
An algebra ${\Bbb A}$ over the reals (${\Bbb R}$) is a composition algebra if it
possesses a unit (denoted as ${\bf 1_{\Bbb A}}$) and a non-degenerate quadratic
form $N$ satisfying
\begin{eqnarray}
N({\bf 1}_{\Bbb A}) &=& 1,\nonumber \\
N(xy) &=& N(x)N(y), \qquad \forall x,y\in {\Bbb A}.
\end{eqnarray}
A composition algebra is alternative if the following left and right alternative properties
are satisfied \cite{schafer}
\begin{eqnarray}
(x^2) y &=& x(xy),\nonumber\\
y x^2 &=& (yx) x.
\end{eqnarray}
A positive definite quadratic form (norm) is a mapping $N: {\Bbb A} \rightarrow {\Bbb R}^+$ s.t. 
\begin{eqnarray}
N(x)=0 &\Leftrightarrow & x=0.
\end{eqnarray}
A composition algebra with positive quadratic form is a division algebra, satisfying
the property
\begin{eqnarray}\label{vee}
xy=0 &\Rightarrow& x=0 \vee y=0.
\end{eqnarray}
Due to the Hurwitz's theorem, the only division algebras are ${\Bbb R}$, ${\Bbb C}$, ${\Bbb H}$
and ${\Bbb O}$.\par
A $\ast$-algebra possesses a conjugation (i.e. an involutive automorphism 
${\Bbb A} \rightarrow {\Bbb A}$) s.t., denoted as $x^\ast$ the conjugate of $x\in {\Bbb A}$, we have
\begin{eqnarray}\label{conjug}
(x^\ast)^\ast &=& x, \nonumber\\
(xy)^\ast &=& y^\ast x^\ast.
\end{eqnarray}
The norm $N(x)$ of an element of a division algebra is expressed as
\begin{eqnarray}
N(x) &=& x x^\ast.
\end{eqnarray}
Besides division algebras, we can introduce their split forms \cite{mcC} as a new set of algebras.
The split-division algebras are $\ast$-algebras with unit. The quadratic form $N$ is
no longer positive-definite and the property (\ref{vee}) is no longer valid.  
The algebras of split complex numbers, split quaternions and split octonions are respectively denoted 
as ${\widetilde {\Bbb C}}$, ${\widetilde {\Bbb H}}$ and ${\widetilde {\Bbb O}}$. The total
number of inequivalent (split)-division algebras over ${\Bbb R}$ is $7$ (the $4$ division algebras and their
$3$ split forms above).
\par
(Split)-division algebras find a unified description through the Cayley-Dickson doubling construction. Given an algebra ${\Bbb A}$ over ${\Bbb R}$, possessing a 
`` $\cdot$ " multiplication, a `` $\ast$" conjugation and a quadratic form $N$, the Cayley-Dickson doubled algebra
${\Bbb A}^2$ over ${\Bbb R}$ is defined in terms of the operations in ${\Bbb A}$. The multiplication, the conjugation and
the quadratic form in ${\Bbb A}^2$ are respectively given by\par~\par
\begin{tabular}{lll}
{\em i}) & multiplication in ${\Bbb A}^2$:&
$(x,y)\cdot (z,w) = (x z+\varepsilon w^\ast y, wx +yz^\ast)$,\\
{\em ii}) &conjugation in ${\Bbb A}^2$:& $(x,y)^\ast = (x^\ast, -y)$,\\
{\em iii})& norm in ${\Bbb A}^2$:&
$N(x,y) = N(x)-\varepsilon N(y)$.
\end{tabular}\par~\\
The unit element ${\bf 1}_{{\Bbb A}^2}$ of ${\Bbb A}^2$ is represented by
${\bf 1}_{{\Bbb A}^2}=({\bf 1}_{{\Bbb A}},0)$.
\\
In the above formulas $\varepsilon$ is just a sign ($\varepsilon=\pm 1$). \par
It is convenient
to denote the Cayley-Dickson's double of an algebra ${\Bbb A}$ by writing the $\varepsilon$ sign on the right
of the original algebra. For division algebras $\varepsilon$ is always negative
($\varepsilon =-1$). We can therefore write\par
${\Bbb C} = {\Bbb R}-$,\par ${\Bbb H}= {\Bbb C}- = {\Bbb R}--$,\par 
${\Bbb O}= {\Bbb H}- = {\Bbb C}--= {\Bbb R}--- $.\\
The split division algebras are obtained by taking a positive ($\varepsilon=+1$) sign.
We have\par
${\widetilde {\Bbb C}} = {\Bbb R}+$,\par
${\widetilde {\Bbb H}} = {\Bbb C}+= \Bbb R-+$ \par
${\widetilde {\Bbb O}} = {\Bbb H}+= \Bbb C-+= \Bbb R--+$. \par
Other choices of the sign produce, at the end, isomorphic algebras. We can, e.g., also write
${\widetilde {\Bbb H}}= {\Bbb R}++$, as well as ${\widetilde {\Bbb O}}= {\Bbb R}+++$.\par
All (split-)division algebras over ${\Bbb R}$ are obtained by iteratively
applying the Cayley-Dickson's construction starting from ${\Bbb R}$.

{}~
\\{}~
\par {\large{\bf Acknowledgments}}{} ~\\{}~\par
This work received support from CNPq (Edital Universal 19/2004). Z. K. acknowledges
FAPEMIG for financial support and CBPF for hospitality.

\end{document}